\documentclass[twocolumn,SC]{revtex4}
\usepackage{graphicx,float}
\usepackage{bm}
\usepackage{amsfonts}
\usepackage{mathrsfs}
\begin{document}
\title{\bf Revisiting YH$_9$ Superconductivity and Predicting High-T$_c$ in GdYH$_5$}
\author{M. A. Rastkhadiv}
\email{rastkhadiv@shirazu.ac.ir}
\affiliation {Estahban Higher Education Center-Shiraz University, Estahban, Iran}

\begin{abstract}
The discovery of superconductivity in $\mathrm{YH_{9}}$ with a critical temperature of approximately $T_c\sim 243 \ K$ has opened a new window toward room temperature superconductivity. In this work, we employ the lowest order constrained variational method to investigate the thermodynamic and magnetic properties of the $\mathrm{YH_{9}}$ structure, obtaining results in good agreement with experimental data.
Based on the robustness of the LOCV approach for describing high-$T_c$ superconductors, we further extend our analysis to the gadolinium-yttrium-hydrogen system across various stoichiometries. The key finding of this study is the prediction of a superconducting phase transition at $T_c = 223.2~\mathrm{K}$ for $\mathrm{GdYH_{5}}$ under a critical pressure of approximately $157~\mathrm{GPa}$. This compound crystallizes in a tetragonal structure with space group $P4/mmm$. Moreover, the calculated gap ratio confirms that $\mathrm{GdYH_{5}}$ is a type-II superconductor with a critical current density suitable for potential industrial applications.

\end{abstract}

\maketitle
\section{INTRODUCTION}

The pursuit of superconductivity at or near room temperature has been one of the most persistent and ambitious challenges in condensed matter physics and materials science. Conventional Bardeen-Cooper-Schrieffer (BCS) theory and its Migdal-Eliashberg extension suggest that extremely high transition temperatures ($T_c$) could in principle be possible if materials with strong electron-phonon coupling and high phonon frequencies can be realized \cite{D1,D2,D3}. For decades, however, the highest $T_c$ among conventional superconductors was restricted to $39\ \mathrm{K}$ in MgB$_2$ \cite{D4}, while unconventional cuprate superconductors reached $T_c$ values of about $133\ \mathrm{K}$ at ambient pressure and up to $\sim$ $164\ \mathrm{K}$ under compression, but without a complete microscopic theoretical understanding \cite{D5,D6}. This limitation motivated new directions in the search for high-$T_c$ materials.

A major conceptual breakthrough came from Ashcrof's idea of chemical 'precompression', where hydrogen, the lightest element with the highest phonon frequencies, could achieve metallization and superconductivity at reduced pressures when combined with heavier atoms \cite{B5}. Although metallic hydrogen itself requires pressures of about $500\ \mathrm{GPa}$ to enter a superconducting state \cite{B4}, hydrogen-rich compounds emerged as more accessible candidates. H$_3$S is a decisive discovery in this regard, which indicates a $T_c$ of $\sim$ $203\ \mathrm{K}$ at $\sim$ $150\ \mathrm{GPa}$, setting a record that exceeded all previously known systems \cite{C1}. This finding, supported by theoretical predictions for the Im-3m phase \cite{B7,B8}, provided the first strong evidence that hydrogen-dominant compounds could serve as high-temperature phonon-mediated superconductors.

Experimental validation of H$_3$S superconductivity was achieved through several complementary techniques, including resistance measurements showing a zero-resistive state, magnetic susceptibility experiments revealing the Meissner effect \cite{C1,C10,C11}, and spectroscopic investigations of the superconducting gap \cite{C12}. Isotope substitution further confirmed the phononic origin of pairing, with large shifts in $T_c$ observed between hydrogenated and deuterated samples \cite{C1,B9,B91,B92,B93,B94,B95,B96,B13}. However, synthesis challenges, including the presence of byproducts such as S, H$_2$S, or H$_4$S$_3$ at high pressures, raised questions about purity and phase stability in different experiments \cite{B8,B11,B13}. Direct synthesis routes from elemental S and H$_2$ later helped to eliminate hydrogen-depleted phases and confirmed the intrinsic superconductivity of Im-3m H$_3$S with a $T_c$ near $200\ \mathrm{K}$ at $\sim$ $150\ \mathrm{GPa}$ \cite{B14,B15,B16}.

The discovery of LaH$_{10}$ marked the next major advance, with superconducting transitions reported near $250-260\ \mathrm{K}$ at pressures of $170-180\ \mathrm{GPa}$ \cite{A8,C20,C21}, in agreement with predictions for the sodalite-like clathrate Fm-3m phase \cite{A4,B5}. Other rare-earth and alkaline-earth superhydrides such as YH$_9$ and YH$_6$ also indicated high $T_c$s of $220-243\ \mathrm{K}$ \cite{A11, A12}, further demonstrating the robustness of the hydrogen-clathrate framework as a platform for high-$T_c$ superconductivity. Chemical substitution, such as alloying La with Y or Ce, has proven to be an efficient method for tuning $T_c$ values, sometimes yielding significant enhancements compared to the pure phases \cite{A24,A25,A26,A27}.
More recently, ternary hydrides have entered the stage, with Li$_2$MgH$_{16}$ predicted to become a superconductor at $\sim$ $473\ \mathrm{K}$ under $250\ \mathrm{GPa}$ due to strong electron donation from Li into the hydrogen lattice \cite{A21}. Meanwhile, compounds such as Li$_5$MoH$_{11}$ and BeReH$_9$ have been synthesized experimentally, although they indicate low $T_c$ values due to the formation of H$_2$ or H$_3$ molecular units \cite{A22,A23}. Notably, yttrium hydrides such as YH$_{10}$ are predicted to achieve $T_c$s approaching or even surpassing room temperature at $250-400\ \mathrm{GPa}$ \cite{A4,A8,A9}, making them prime targets for future high-pressure investigations.

Although most theoretical predictions of high-T$_c$ superconductors are based on density functional theory (DFT) combined with Eliashberg theory or the Allen-Dynes modified McMillan equation \cite{allen}, a new formalism is appeared in this field. This formalism is lowest order constrained variational (LOCV) method. It is a many-body method  written for nuclear physics \cite{13}. Bordbar et. al applied this method in condensed matter physics by calculating the thermodynamic properties of liquid $^3$He \cite{e3} by this method.
Another application of this formalism in condensed matter physics concerns the liquefaction of 
$^3$He in strictly two dimensions. Despite theoretical investigations suggesting the absence of 
$^3$He liquefaction in this regime \cite{hel1,hel2}, low-temperature heat-capacity measurements by Sato \textit{et al.} \cite{sat1,sat2} clearly demonstrated a liquid-gas phase transition for fluid $^3$He adsorbed on a graphite surface in two dimensions. To the best of our knowledge, the LOCV method is the only theoretical approach that has successfully confirmed the experimental results of Sato \textit{et al.} for the liquefaction of $^3$He on a graphene sheet \cite{jpn}.
This method has also been successfully applied to computing phase transitions and electronic structural stability in highly correlated systems as well \cite{my,e4}. Very recently, LOCV method has applied in superconductivity by calculating the critical temperatures and various thermodynamic properties of CeH$_9$ structure \cite{e5} with outstanding compatible results with experimental data \cite{wuh}. Ref. \cite{e5} also could predict a high-T$_c$ superconductivity ($199\ \mathrm{K}$) in two corrugated graphene sheets with intercalated CeH$_9$ molecules at ambient pressure which has not been verified experimentally yet. 
The notable agreement between the theoretical calculations of LOCV method and experimental observations shows the efficacy of the LOCV method in accurately identifying thermodynamic properties of highly correlated fermions. 
More details regarding numerous successful achievements of LOCV method in different branches of physics can be found in review paper of Ref.\cite{revp}. 

Based on these investigation, in present work, we apply the LOCV method to investigate the experimental data of Kong \textit{et al.} survey on finding the T$_c$ of superconductivity in YH$_9$ structure at high pressures \cite{A11}, and try to find similar compounds to have high critical points.
As a result, different combinations of Y-H-Gd are considered.
This paper is outlined as follows: In Sec. II, we present 
the LOCV method and explain the configuration of the system. In Sec. III, the thermodynamic properties of the system are presented.
The final section provides a summary and conclusions.

\section{Method}
The LOCV method is applied to calculate the thermodynamic properties of the system. This method does not introduce any free parameters into calculations \cite{13}, therefore, it is appropriate for systems with long correlation length. 
The total energy $(E)$ of the system is divided
into energy clusters $(E=E_1+E_2+...,)$. These terms will be explained later.
The unique feature of this method, compared to other many-body approaches, is its ability to account for all interactions in the system and to compute higher-order energy cluster terms until the results converge within reasonable time and computational cost, while avoiding unnecessary calculations.

\subsection{Lowest-order constrained variational method}
As the system we intend to verify is electrons (N$^{(+)}$ spin-up and N$^{(-)}$ spin-down), the Fermi-Dirac mean occupation number ($ \rho^ {(i)} $)
is defined by,
\begin{equation}
{\rho^{(i)}(k)}=[e^{\beta(\varepsilon(k)-\mu^{(i)})}+1]^{-1},
\end{equation}
where $i=+,-$ is spin projection, $\varepsilon(k)$ is the single particle energy, $k_B$ is the Boltzmann constant, $\mu$ is the chemical potential, $\beta=\frac{1}{k_B T}$, and $k$ is the wave vector of an electron.
For each number density ($\mathfrak{n}$) and temperature ($T$), the chemical potential is determined from the particle number constraint as follows,
\begin{equation}
N=\sum_{k,i}[e^{\beta(\varepsilon(k)-\mu^{(i)})}+1]^{-1}.
\end{equation}
The spin order parameter ($ \xi $) is defined as,
 \begin{equation}\label{eq656}
 \xi=\frac{N^{(+)}-N^{(-)}}{N}. 
\end{equation}

In LOCV formalism, the $N$ particles total wave function is devided into two parts as follows,
\begin{equation}
\psi( 1,2,...,N)=F(1,2,...,N)\Phi( 1,2,...,N)\label{eq1},
\end{equation}
where $ F $ is the $N$ particles correlation function operator and $\Phi$ is the total wave function of the non-interacting particles  \cite{13}. As the system particles are fermions,  $\Phi$ denotes the Slater determinant. The Jastrow approximation is applied as follows,
\begin{equation}
F( 1,2,...,N)= \prod_{i<j} f_{2}(r_{ij}) \; \prod_{i<j<k} f_{3}(r_i,r_j,r_k) \cdots ,
\end{equation}
where $f_{2}(r_{ij})$ and $f_{3}(\mathbf{r}_i, \mathbf{r}_j, \mathbf{r}_k)$ 
are the two-body and three-body correlation functions, respectively. 
The two-body term $f_{2}$ considers pair interactions, while $f_{3}$ incorporates three-particle 
correlations depending on the relative positions of triplets of particles.
Higher-order terms are included until convergence is achieved.

Within the LOCV formalism, the system's total energy is expressed through a cluster expansion of correlated contributions
\begin{equation}\label{eq2}
E=\frac{{\left\langle {\Psi} \right|{ \mathcal{H} }\left| {\Psi} \right\rangle }}{\langle\Psi\mid\Psi\rangle}=E_1+E_2+...,
\end{equation}
where $\mathcal{H}$ is the total Hamiltonian, $E_1$ is the one-body and $E_2$ is the two-body cluster energies, etc. 
$E_1$ and $E_2$ are defined as,
\begin{equation}\label{eq3}
E_1=\sum_{i}E^{(i)}_1=\sum_{k,i}\rho^{(i)}(k)\varepsilon(k),
\end{equation}
and
\begin{equation}\label{eq5}
E_2 = \frac{1}{2}\sum_{i,j,{j_1},{j_2}}\rho^{(i)}(k_{j_1})\rho^{(j)}(k_{j_2}) {\left\langle {{j_1}{j_2}} \right|{\omega(12)}\left| {{j_1}{j_2} } \right\rangle }_a.
\end{equation}
Here, $j_1$ and $j_2$ are the quantum numbers associated with a two-body state, the parameter $a$ denotes the antisymmetric nature of the corresponding wave function, and $\omega(12)$ is the effective interaction potential for the two-body cluster, defined as follows
\begin{equation}\label{eq51}
\omega(12)=\frac{\hbar^2}{m^{\ast}}\left[ \vec{\nabla} F(\mathtt{r} )\right]  ^2+F^2(\mathtt{r} )V(\mathtt{r} ),
\end{equation} 
where $\hbar = h / (2\pi)$ is the reduced Planck's constant, the symbol $ \vec{\nabla}$ represents the gradient operator, $V(\mathtt{r})$ specifies the inter-particle potential as a function of the separation $r$ between particles, and $m^{\ast}$ is the effective mass associated with the valence electrons.
To include the effects of spin–spin interactions, the two-body correlation function is written as follows,
\begin{equation}\label{eq511}
F(\mathtt{r} ) =\sum_{s=0,1}  f_s  (\mathtt{r} )P_s , 
\end{equation}
where,
\begin{equation}\label{eq512}
P_0 =\frac{1}{4}(1 -\vec{\sigma}_1  . \vec{\sigma}_2 ),
\end{equation}
and
\begin{equation}\label{eq513}
P_1 =\frac{1}{4}(3+ \vec{\sigma}_1 . \vec{\sigma}_2 ).
\end{equation}
Here, $f_{0}$ and $f_{1}$ represent the two-body correlation functions in the
spin-singlet and spin-triplet channels, respectively, while $\sigma$ is the
electron spin. By substituting Eq.~(\ref{eq511}) into Eq.~(\ref{eq51}) and
performing the required algebra, one obtains the spin-dependent relation for
$E_{2}$,
\begin{equation}\label{eq514}
E_{2}= \sum_{s=0,1}E_{2,s},
\end{equation}
where,
\begin{equation}\label{eq515}
E_{2,s} =2\pi \mathfrak{n}\int_{0}^{\infty } d\mathtt{r} \, \mathtt{r} ^2 \left[\frac{\hbar^2}{m^{\ast}}\left[ \bigtriangledown f_s(\mathtt{r}) \right]^2+f_s^2(\mathtt{r} )V(\mathtt{r} )\right] a_s .
\end{equation}
In this equation, the quantities $a_0$ and $a_1$ are given by
\begin{eqnarray}\label{eq516}
a_0 =\frac{1}{4}\left[ 1-\xi ^2+\gamma^{(+)}(\mathtt{r} )\gamma^{(-)}(\mathtt{r} ) \right],\nonumber\\
a_1 =\frac{1}{4}\left[ 3+\xi ^2-\gamma^{(+)}(\mathtt{r} )\gamma^{(-)}(\mathtt{r} ) \right. \nonumber\\
-\left. \left[ \gamma^{(+)}(\mathtt{r} )\right]^2-\left[ \gamma^{(-)}(\mathtt{r} ) \right]^2 \right],
\end{eqnarray}
where,
\begin{equation}\label{eq517}
\gamma^{i}(\mathtt{r} )=\frac{1}{\pi^2\mathfrak{n}}\int_{0}^{\infty } dk \, k^2\frac{sin(k\mathtt{r} )}{k\mathtt{r} }\rho^{(i)}(k).
\end{equation}

The minimization of the two-body energy, subject to the normalization condition \cite{13}, is done by varying the two-body correlation function. This procedure leads to the following Euler-Lagrange differential equation,
\begin{equation}\label{eq64}
f_s^{\prime\prime}(\mathtt{r} )+\left( \frac{2}{\mathtt{r} }+\frac{a^{\prime}_s}{a_s}\right) f^{\prime}_s(\mathtt{r} )
-\frac{m^{\ast}}{\hbar^2} \left( V(\mathtt{r} )-2\Lambda \right) f_s(\mathtt{r} )=0.
\end{equation}
The Lagrange multiplier $\Lambda$ imposes the normalization condition, and the prime notation indicates differentiation with respect to $\mathtt{r}$. The spin-dependent pair correlation function is obtained by solving Eq.~(\ref{eq64}), from which the energy clusters are subsequently evaluated. Since the contribution of higher-order clusters to the total system energy decreases progressively, the cluster expansion is carried out only up to the point where additional terms no longer make a significant impact on the overall energy.
Additional information on the higher orders of of energy clusters can be found in Ref.~\cite{13}.

\subsection{Magnetic susceptibility}
The magnetic susceptibility ($\chi$) is a fundamental quantity for analyzing the magnetic response of materials. 
In this regard, we assume an external magnetic field oriented along the $z$-axis ($H_z$) and add the 
interaction term $-\sum_{i}\vec{\mu}_i \cdot \vec{H}$ into the system Hamiltonian. 
Here, the magnetic moment of the $i$-th electron is defined as 
$\vec{\mu}_i = -\frac{e}{m^{\ast}}\vec{\sigma}_i$, where $e$ is the electron charge.
The magnetic susceptibility can be derived as follows,
\begin{equation}
M_z=\left( \frac{\partial \mathcal{F} }{\partial H_z }\right)_{T,n} ,
\end{equation}
\begin{equation}
\chi=\left(\frac{\partial H_z}{\partial M_z}\right)_{T,n},
\end{equation}
where $M_z$ is the magnetization along the $z$–axis, and $\mathcal{F}$ represents the free energy associated with the valence electrons system. 

 \subsection{System Setup}

As will be discussed in subsequent section, the crystal structures are investigated in this work include 
$\mathrm{YH}_{9}$, $\mathrm{GdYH}_{5}$, $\mathrm{GdYH}_{6}$, $\mathrm{GdYH}_{8}$, and $\mathrm{GdYH}_{10}$. Since the computational methodology employed to evaluate their thermodynamic 
properties is identical for all cases, we only present the procedure explicitly for $\mathrm{YH}_{9}$.
The calculation of thermodynamic quantities requires knowledge of the equilibrium atomic coordinates. 
It should be emphasized that these equilibrium positions correspond to the quantum-mechanical 
expectation values of the atomic positions, representing probability-density averaged configurations. 
As an initial step, trial atomic positions for a $\mathrm{YH}_{9}$ molecule is established based on structural 
considerations. These trial coordinates are then systematically refined through a structural 
optimization procedure to give the equilibrium configuration of the molecule.
To determine the trial coordinates of $\mathrm{YH}_{9}$ molecules, we first address the spatial configuration of an individual $\mathrm{YH}_{9}$ unit. The coordinates obtained for this isolated molecule are then adopted as the initial (trial) coordinates for all $\mathrm{YH}_{9}$ molecules within the periodic lattice.
To model the spatial domain of hydrogen-yttrium interactions in a $\mathrm{YH}_{9}$ molecule, we define an idealized spherical region of radius approximately $3.2 \ \text{\AA}$ centered on the $\mathrm{Y}$ atom. This sphere indicates the effective zone in which hydrogen atoms can interact both with the central $\mathrm{Y}$ atom and with one another. The method used to determine this effective radius is described in detail in the Results and Discussion section. For computational purposes, the spherical region is discretized into cubic subspaces, each with a volume of $10^{-3} \ \text{\AA}^3$.
The free energy of a single $\mathrm{YH}_{9}$ molecule is evaluated using the LOCV method. This calculation is performed for all possible effective configurations arising from the distribution of nine hydrogen atoms among the cubic subspaces, with each hydrogen atom positioned at the center of a distinct cube. The equilibrium arrangement of the nine hydrogen atoms surrounding the central $\mathrm{Y}$ atom is identified as the configuration corresponding to the minimum free energy.
All interatomic interactions are modeled through Coulombic potentials, supplemented by spin-spin interactions as defined in Eq. (\ref{eq515}). The computational approach for determining the free energy of a single $\mathrm{YH}_{9}$ molecule using the LOCV framework parallels that employed for evaluating the free energy of valence electrons, as detailed in Eqs. (\ref{eq10}), (\ref{eq11}), and (\ref{eq12}).

These accurate coordinates of a $\mathrm{YH}_{9}$ molecule are applied only as initial approximations in lattice computations. The true equilibrium positions of the $\mathrm{Y}$ and $\mathrm{H}$ atoms in the periodic structure are subsequently determined through a variational procedure. In this regard, the preliminary positions of the $\mathrm{Y}$ and $\mathrm{H}$ atoms are denoted by vector $\mathbf{R}^\prime_Y$ and $\mathbf{R}^\prime_{H}$, respectively. Using these trial (initial) coordinates, the free energy of the entire lattice structure is computed by the LOCV method.
In the second stage, the atoms are allowed to relax and attain their most stable configurations, analogous to the positions they naturally adopt in experimental conditions. Through this procedure, the theoretical model is adjusted to better correspond with experimental observations. The atomic degrees of freedom are introduced by adding a small displacement term, $ \delta\mathbf{R}^\prime $, to the trial coordinates such that
$\mathbf{R}_Y = \mathbf{R}^\prime_Y + \delta\mathbf{R}^\prime_Y$ and
$\mathbf{R}_H = \mathbf{R}^\prime_H + \delta\mathbf{R}^\prime_H$.
The free energy of the lattice is then minimized with respect to variations in $ \delta\mathbf{R}^\prime_Y $ and $ \delta\mathbf{R}^\prime_H $, giving the optimized atomic positions. Consequently, the vectors $\mathbf{R}$ indicate the true equilibrium coordinates of the atoms within the lattice.

The electrons associated with both atoms are categorized into valence and non-valence groups. The valence electrons constitute the thermodynamic system whose critical behavior is the primary focus of this study. In this framework, these electrons are not localized to individual atoms but instead move throughout the crystal lattice, forming a fluid that gives rise to electrical conductivity or superconductivity. Accordingly, we refer to them as fluid electrons.
In contrast, the non-valence electrons remain bound to their respective nuclei, producing a shielding effect that inhibits their participation in conductive or superconductive processes. Within our calculations, each yttrium and hydrogen atom is treated as an ion carrying an effective charge $q$, around which the fluid electrons are distributed. Specifically, the ions are represented as Y$^{3+}$ and H$^{+}$, devoid of fluid electrons. The effective charge $q$ of each ion is computed by the positive nuclear charge and the screening influence of the non-valence electrons.

In the LOCV framework, the single-particle energy $\varepsilon(k)$, the single-particle wave function $\Phi(k)$, and the interparticle potentials are fundamental input parameters. Once these quantities are determined, all relevant thermodynamic properties can be evaluated. To obtain $\varepsilon(k)$ and $\Phi(k)$, the following
 Schr\"{o}dinger equation is solved for a fluid electron located at $\mathbf{r} = x\hat{i} + y\hat{j} + z\hat{k}$, interacting with the electrostatic potential caused by the Y$^{3+}$ and H$^{+}$ ions.
\begin{equation}\label{eq10}
-\frac{\hbar^2\nabla^2}{2m^{\ast}} \Phi +U \Phi=\varepsilon \Phi, 
\end{equation}
where,
\begin{equation}\label{eq11}
U=\sum_{i=1}^{N_1} V_{ e-Y^{3+}}(\mathbf{R}_{Y_i},\mathbf{r}) + \sum_{j=1}^{N_2} V_{ e-H^+}(\mathbf{R}_{H_j},\mathbf{r}).
\end{equation}
$ \nabla^2 $ is the Laplacian operator, while $ N_1 $ and $ N_2 $ represent the maximum numbers of yttrium and hydrogen ions, respectively, that interact effectively with a fluid electron. These effective coordination numbers are determined by identifying the distances at which the electron-yttrium (e-Y$^{3+}$) and electron-hydrogen (e-H$^{+}$) interaction potentials diminish by two orders of magnitude from their maximum absorbing absolute values.

\begin{figure*}[t] 
 \centering
 \includegraphics[scale=0.58]{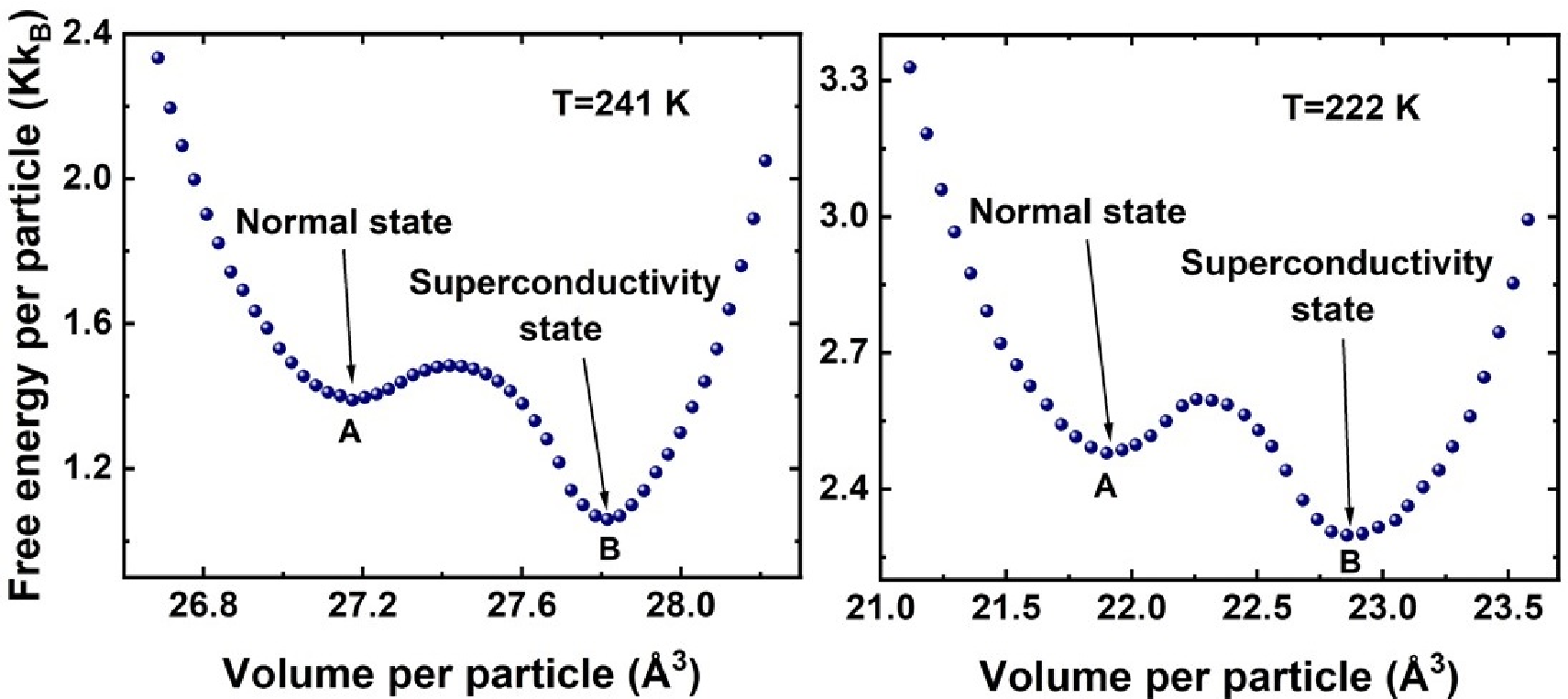}
 \caption{ Dependence of the free energy per electron on the volume per particle for $\mathrm{YH}_{9}$ (left) $\mathrm{GdYH}_{5}$ (right). Points A correspond to the normal electronic state, while points B indicate a superconducting state for the fluid electrons. Points B represent a more stable state, as they correspond to a lower free energy.}
  \label{fig:wide}
\end{figure*}

A fluid electron is subject to three distinct types of interactions: (1) electron-yttrium ions (e-Y$^{3+}$), (2) electron-hydrogen ions (e-H$^{+}$), and (3) electron-electron (e-e) potentials. The e-Y and e-H interactions are accounted for in Eq. (\ref{eq11}), whereas the e-e interaction is incorporated through Eq. (\ref{eq5}) and in the higher-energy clusters. The general solution to Eq. (\ref{eq10}) is expressed as follows:
\begin{equation}\label{eq12}
\Phi_n({\mathbf{r},\mathbf{k}})=u_n(\mathbf{r},\mathbf{k})e^{i\mathbf{r}.\mathbf{k} }.
\end{equation}
The choice of this wave function is motivated by the inherent periodicity of the crystal lattice. Owing to these periodicities, the wave function is expressed in Bloch form \cite{ash}, where the function $u_n(\mathbf{r},\mathbf{k})$ satisfies the Born-von Karman periodic boundary condition. Considering these boundary conditions into Eq.~(\ref{eq10}), the  Schr\"{o}dinger equation is solved using the shooting method, giving the eigenvalues $\varepsilon(k)$ and the corresponding eigenfunctions $\Phi(k)$. The total energy of the fluid electron is subsequently evaluated by substituting $\varepsilon(k)$ and $\Phi(k)$ into the energy clusters and summing their contributions.

\section{Results and Discussions }\label{NLmatchingFFtex}

The first aim of the present study is to apply the LOCV method to revisit the experimental data of Ref. \cite{A11} and compute additional thermodynamic and magnetic properties of the system. The second aim is to investigate Y-H structures combined with Gd in search of high-$T_c$ superconductivity. Several Y-H-Gd compositions with different stoichiometries were considered. Based on free-energy calculations, only $\mathrm{GdYH}_{5}$, $\mathrm{GdYH}_{6}$, $\mathrm{GdYH}_{8}$, and $\mathrm{GdYH}_{10}$ indicate stable states at $P_c \sim 100 \ GPa$. As our focus is restricted to superconducting phases with $T_c$ above $200\ \mathrm{K}$, and no superconducting transition was observed for these compounds except for $\mathrm{GdYH}_{5}$, we report only the results corresponding to $\mathrm{GdYH}_{5}$.

The energy clusters of the fluid electrons are computed up to the $E_4$ term. Since the results converge at this order, higher-order contributions are neglected. The total energy per electron of the system is then obtained by summing these cluster terms.
It is more convenient to perform the analysis within the canonical ensemble; therefore, the free energy function is employed in the subsequent calculations. The free energy is minimized with respect to $m^{\ast}$, $\delta\mathbf{R}^\prime_{\mathrm{Y}}$, $\delta\mathbf{R} ^\prime_{\mathrm{Gd}}$, and $\delta\mathbf{R}^\prime_{\mathrm{H}}$ over all temperatures and densities, a procedure referred to as structural relaxation.
The results indicate a pronounced increase in the free energy when the Y-H and Gd-H separation exceeds approximately $3.2\ \text{\AA}$ and $2.9\ \text{\AA}$, respectively. Accordingly, the minimization process is restricted to interatomic distances shorter than $3.2\ \text{\AA}$ $2.9\ \text{\AA}$, respectively.
 
Plotting the free energy diagram is a fundamental method for identifying phase transitions. 
Figure~\ref{fig:wide} displays the free energy per electron as a function of 
$\mathfrak{n^{-1}}$ for $\mathrm{YH}_{9}$ at $T = 241~\mathrm{K}$ and for $\mathrm{GdYH}_{5}$ at $T = 222~\mathrm{K}$, respectively. The critical pressures are $P_c = 201.1~\mathrm{GPa}$ and $P_c = 157.8~\mathrm{GPa}$, respectively.
The presence of two minima in both figures implies the occurrence of a phase transition. 
As will be shown later, the specific heat curves for both compounds diverge at the transition points, 
confirming that these are second-order phase transitions. 
These two minimum points appear\begin{scriptsize}\end{scriptsize} at temperatures $T \leq T_c = 241.3~\mathrm{K}$ for $\mathrm{YH}_{9}$ and 
$T \leq T_c = 223.2~\mathrm{K}$ for $\mathrm{GdYH}_{5}$. 
Above these critical temperatures, the two minima merge into a single minimum.
It should be emphasized that within a variational framework such as the one employed in this work, 
only the relative minima are physically stable phases. 
In Fig.~\ref{fig:wide}, points A and B, located at 
$\mathfrak{n^{-1}} = 1.389~\text{\AA}^{3}$ and $\mathfrak{n^{-1}} = 1.06~\text{\AA}^{3}$ for $\mathrm{YH}_{9}$, and $\mathfrak{n^{-1}} = 2.479~\text{\AA}^{3}$ and $\mathfrak{n^{-1}} = 2.298~\text{\AA}^{3}$ for $\mathrm{GdYH}_{5}$.
These points represent the normal and superconducting states, respectively.

\begin{figure}
\includegraphics[scale=.36]{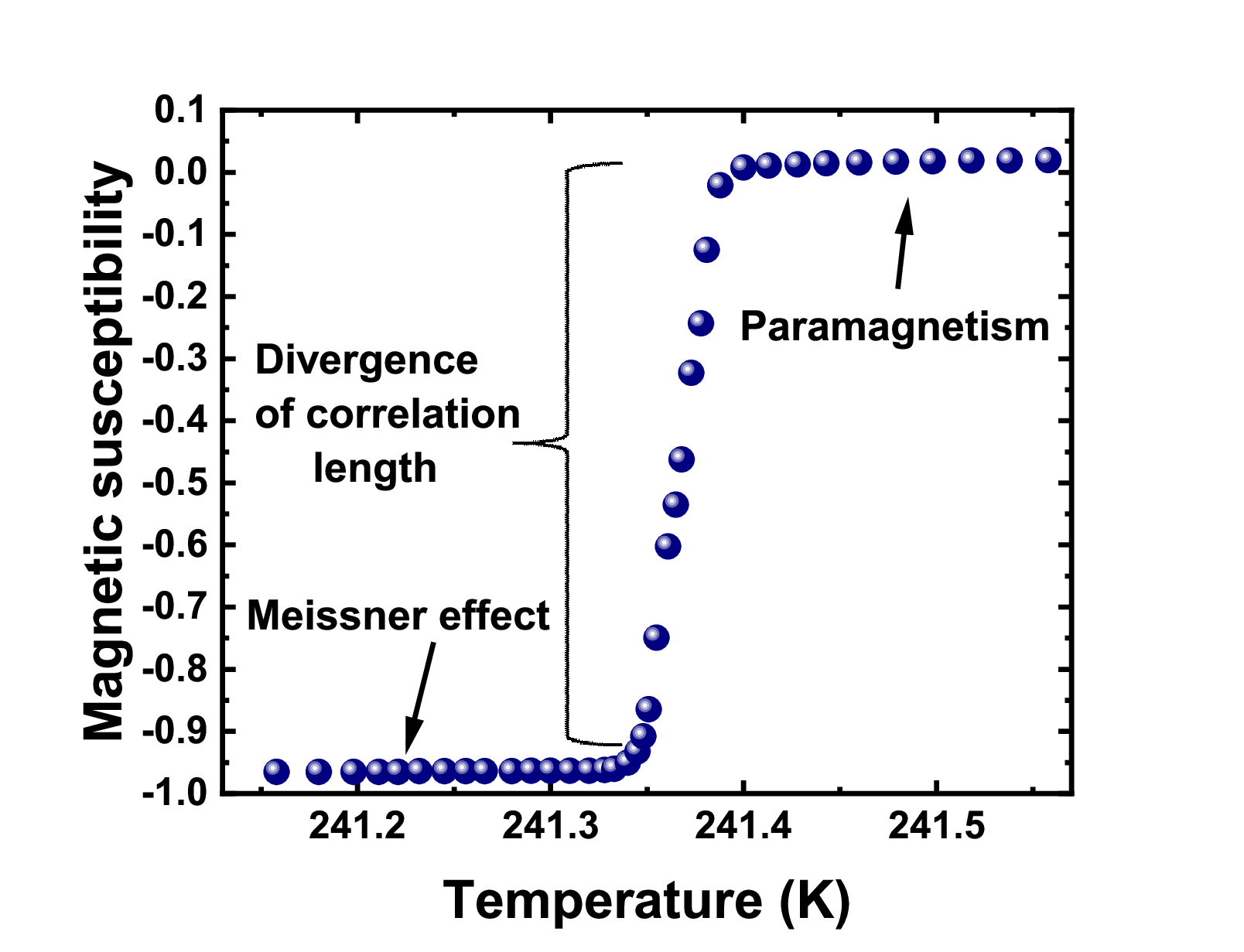}
\caption{ 
The magnetic susceptibility of the fluid electrons in $\mathrm{YH}_{9}$ near the phase transition is shown. It is small and positive above $T_c$, indicating a paramagnetic response. For $T \lesssim T_c$, the correlation length diverges due to the second-order phase transition, causing $\chi$ to rapidly decrease to approximately $-1$, which indicates the appearance of the Meissner effect.
}\label{susceptibility}
\end{figure}
\subsection{Criticality in fluid electrons}

The superconducting transition temperature in phonon-mediated high-pressure superconductors is typically estimated by evaluating the Eliashberg spectral function, $\alpha^2F(\omega)$, through density functional theory or density functional perturbation theory. From $\alpha^2F(\omega)$, the electron-phonon coupling constant ($\lambda$) and the logarithmic average phonon frequency ($\omega_{\log}$) are determined and inserted into the Allen-Dynes modified McMillan equation to predict $T_c$. 
In contrast, within the LOCV method, the transition temperature $T_c$ is obtained directly from the free energy, providing a more fundamental description of the superconducting phase transition.
A distinctive feature of this study is the use of a single wave function to describe both the normal and superconducting regimes, without any change to the underlying Hamiltonian. In this formulation, the phase transition and the spontaneous symmetry breaking emerge naturally from the collective dynamics of the electron fluid. This unified treatment across the critical temperature offers a coherent picture of how electronic correlations drive the transition.

\begin{figure}
\includegraphics[scale=0.355]{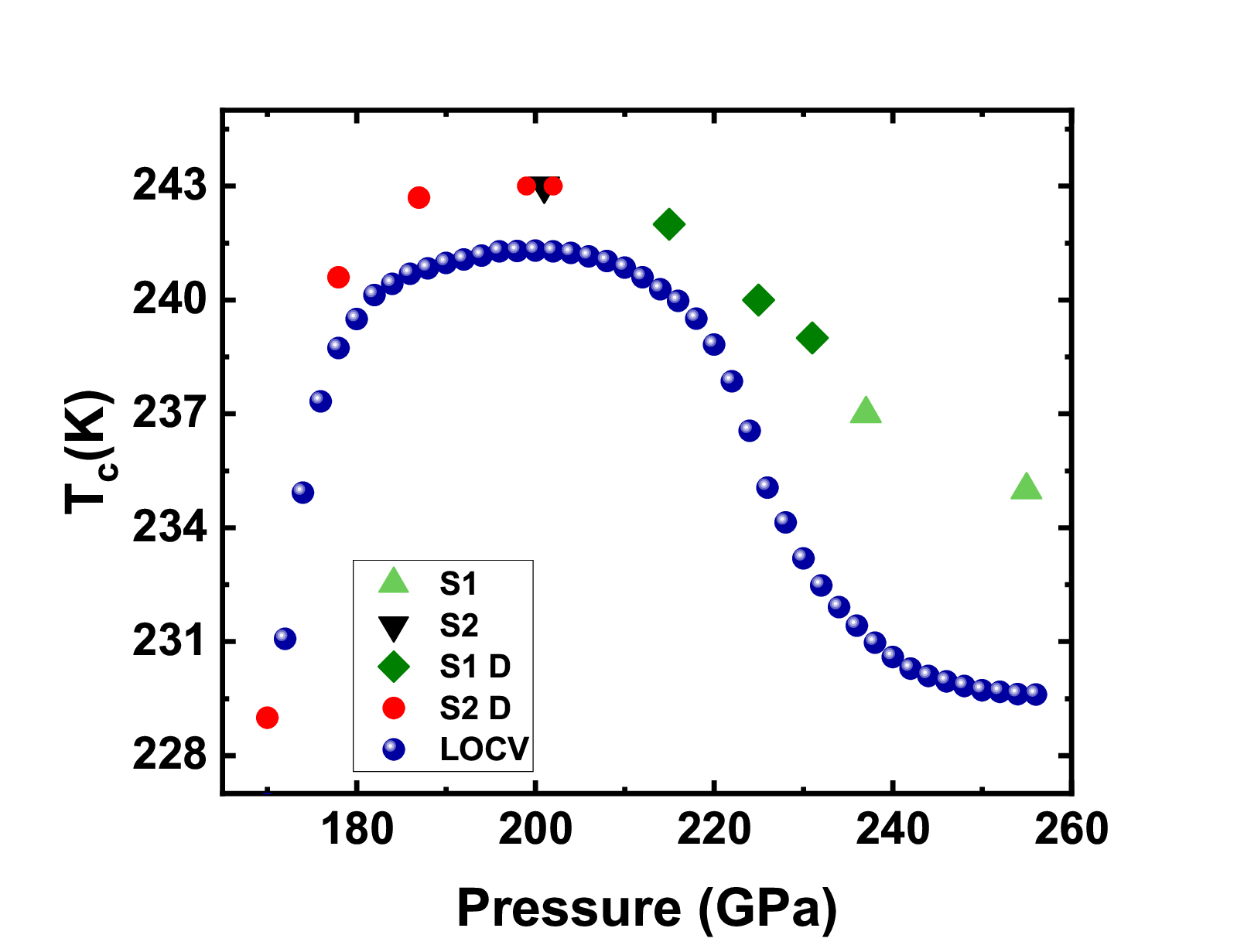}
\caption{
$T_c$ as a function of pressure for superconductivity in $P6_{3}/mmc$ $\mathrm{YH}_{9}$. The labels $\mathrm{S}$ and $\mathrm{D}$ represent “sample” and “decreasing pressure,” respectively. Experimental data are taken from Ref.~\cite{A11}; further details about the samples are provided therein.
}\label{tcpressure}
\end{figure}

Our calculations show that the effective mass lies within the range 
\( 0.12 \leq \frac{m^{\ast}}{\mathfrak{m}} \leq 0.32 \) for 
\(\mathrm{YH}_{9}\) and 
\( 0.09 \leq \frac{m^{\ast}}{\mathfrak{m}} \leq 0.41 \) for 
\(\mathrm{GdYH}_{5}\), 
where \( \mathfrak{m} \) denotes the free-electron mass.
The reason that LOCV is able to compute the spontaneous symmetry breaking leading to Figs.~\ref{fig:wide} is governed by two principal factors: the consideration of the effective mass $m^{\ast}$ as a variational parameter and the incorporation of the displacement variational parameter $\delta \mathbf{R}^\prime$ in the formalism.
The omission of any of these parameters suppresses the second-order phase transition. The large number of hydrogen atoms is particularly significant, as it gives rise to high-frequency phonon spectra that promote Cooper-pair formation~\cite{ash2, cohen}.
For $\mathrm{YH}_{9}$, the variational displacements are found within the ranges
$1.38 \le \delta\vec{R}^\prime_Y \le 1.64~\text{pm}$ and
$1.92 \le \delta\vec{R}^\prime_H \le 2.07~\text{pm}$.
In the case of $\mathrm{GdYH}_{5}$, the corresponding ranges are
$0.81 \le \delta\vec{R}^\prime_{\mathrm{Gd}} \le 1.12~\text{pm}$,
$1.22 \le \delta\vec{R}^\prime_{Y} \le 1.74~\text{\AA}$, and
$1.98 \le \delta\vec{R}^\prime_H \le 2.20~\text{\AA}$.
As anticipated, the heavier atoms indicate lower mobility, whereas the hydrogen ions display enhanced mobility due to their low atomic mass and high tendency to form bonds. 
Notably, the present formalism consider these displacements through the minimization of the total free energy of the ionic lattice. 
The calculated mobilities of Gd, Y, and H ions are relatively small, indicating that their trial positions remain nearly unchanged. Larger ionic movements would significantly alter the lattice configuration and complicate the solution of the Schr\"{o}dinger equation.

\begin{figure}
\includegraphics[scale=0.32]{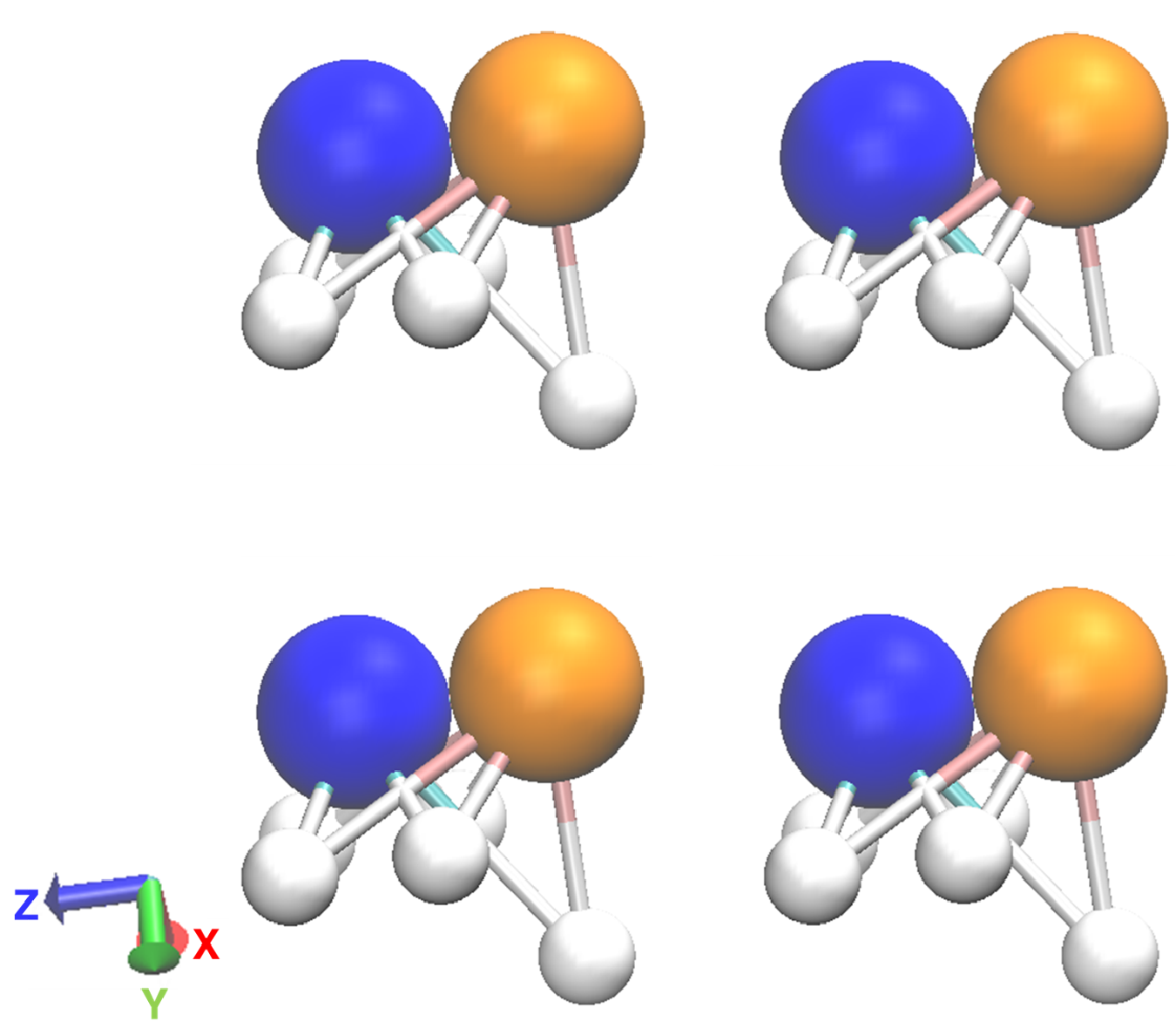}
\caption{
Crystal structure of tetragonal $\mathrm{GdYH_{5}}$ with space group $P4/mmm$. Blue, orange, and white spheres represent Gd, Y, and H atoms, respectively. Atomic sizes are adjusted for clarity, and the distances between molecular units are increased to provide a clearer perspective.
}\label{structure}
\end{figure}

A characteristic property of superconductivity is the Meissner effect. To verify this phenomenon, the magnetic susceptibility ($\chi$) of $\mathrm{YH}_{9}$ is plotted in Fig.~\ref{susceptibility} near the superconducting transition. The LOCV calculations are performed under an external magnetic field applied along the $z$-axis, and the system's magnetic response was evaluated accordingly. 
At temperatures above the critical temperature ($T > T_c$), $\chi$ demonstrates a weakly positive (paramagnetic) response. As the temperature approaches $T_c$, the correlation length diverges, resulting in a rapid decrease of $\chi$ to nearly $-1$ just below the transition; for example, $\chi = -0.96$ at $T = 241.3~\mathrm{K}$. This sharp change in susceptibility provides clear evidence of a second-order phase transition from the normal (paramagnetic) state to the superconducting (diamagnetic) state.
The ideal Meissner state corresponds to $\chi = -1$. The slight deviation observed in Fig.~\ref{susceptibility} arises from the approximations in the LOCV treatment, particularly in the construction of the correlated many-body wave function and the evaluation of the cluster energy contributions.

To assess the consistency of the LOCV calculations with experimental observations, the LOCV-derived phase diagram of the $P6_3/mmc$ $\mathrm{YH}_{9}$ structure is presented in Fig.~\ref{tcpressure}. For comparison, the experimental data from Ref.~\cite{A11} are also included.
S1 and S2 are two samples investigated in Ref.~\cite{A11}, and the D symbol indicates data obtained during subsequent decompression.
 At lower pressures, the theoretical predictions indicate excellent agreement with the experimental results, while a slight deviation appears at higher pressures. The LOCV method successfully reproduces the curvature evolution of the phase diagram, in good correspondence with experiment. In particular, an inflection point emerges near $P = 225~\mathrm{GPa}$ in the LOCV results, consistent with the experimentally observed feature.
Furthermore, as shown in Fig.~\ref{tcpressure}, the superconducting transition temperature of $\mathrm{YH}_{9}$ demonstrates a strong pressure dependence. At pressures below approximately $P = 180~\mathrm{GPa}$, $T_c$ decreases rapidly with reducing pressure.

\begin{figure}
\includegraphics[scale=0.36]{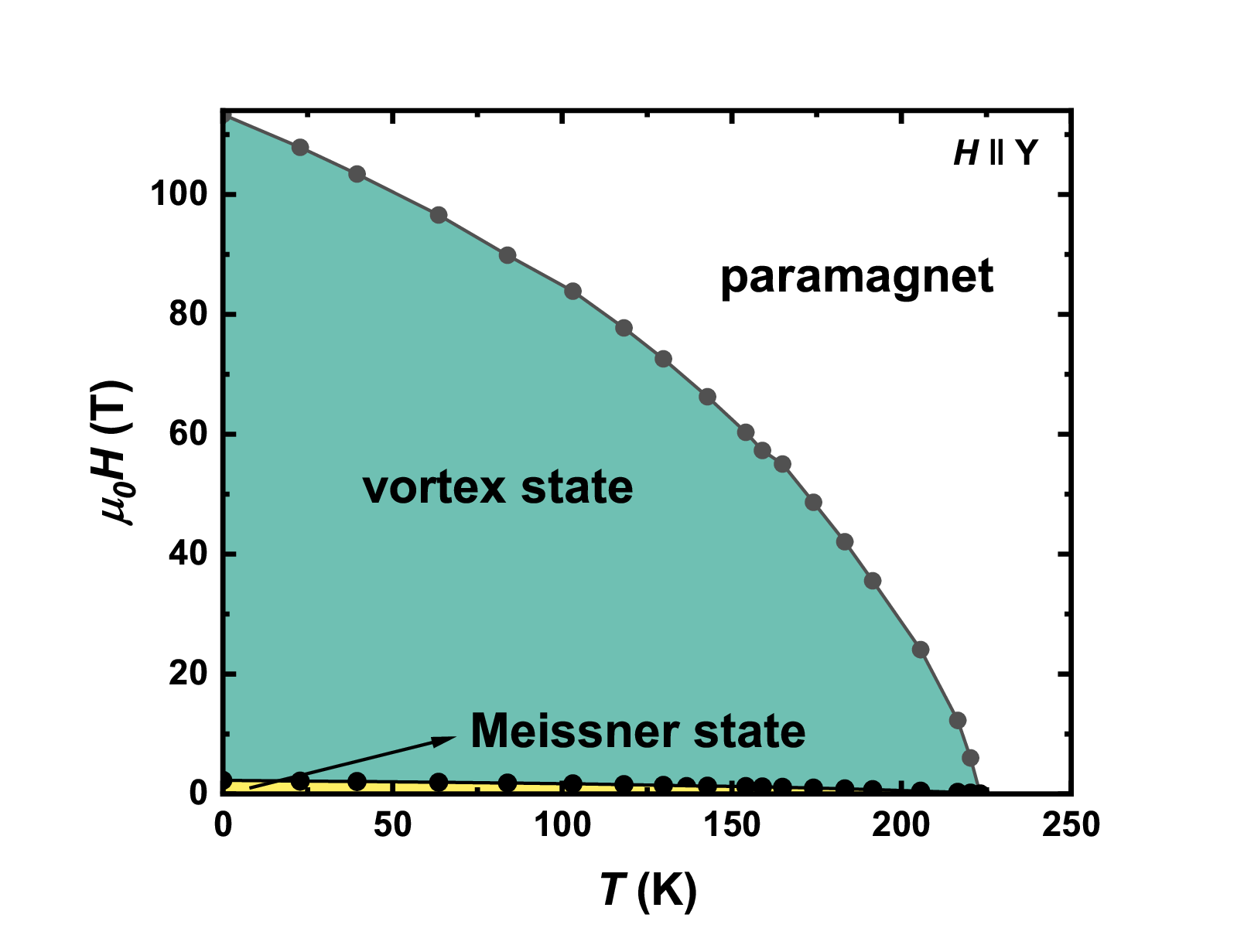}
\caption{
Phase diagram of $\mathrm{GdYH_{5}}$ illustrating the perfect diamagnetic (Meissner) state below $H_{c_1}=2.26~\mathrm{T}$, the mixed (vortex) state between $H_{c_1}$ and $H_{c_2}=113.34~\mathrm{T}$, and the paramagnetic phase above $H_{c_2}$. The magnetic field is applied along the Y direction.
}\label{phase}
\end{figure} 

The free energy of $\mathrm{GdYH}_{5}$ is calculated up to a pressure of $280~\mathrm{GPa}$. The maximum superconducting transition temperature, $T_c = 223.2~\mathrm{K}$, occurs at $188.46~\mathrm{GPa}$. By minimizing the free energy with respect to the lattice parameters at this pressure, the equilibrium lattice structure of $\mathrm{GdYH}_{5}$ is obtained, as illustrated in Fig.~\ref{structure}.
This structure crystallizes in the tetragonal crystal system with space group $P4/mmm$.
 The Gd-Y bond length is $1.12~\text{\AA}$, while the Gd-H and Y-H bond lengths range from $1.18$ to $1.78~\text{\AA}$ and from $1.16$ to $1.69~\text{\AA}$, respectively.
The $H$-$T$ phase diagram of $\mathrm{GdYH}_{5}$ at $188.46~\mathrm{GPa}$ for an applied magnetic field H$ \parallel $Y is shown in Fig.~\ref{phase}. The yellow, green, and white regions correspond to the Meissner, vortex, and paramagnetic states, respectively. The lower and upper critical fields are $H_{c_1} = 2.26~\mathrm{T}$ and $H_{c_2} = 113.34~\mathrm{T}$. Similar to other superhydride superconductors, $\mathrm{GdYH}_{5}$ indicates a remarkably high $H_{c_2}$ value, which can be attributed to strong electron-phonon coupling associated with the high-frequency phonon modes of hydrogen.

\begin{figure}
\includegraphics[scale=0.32]{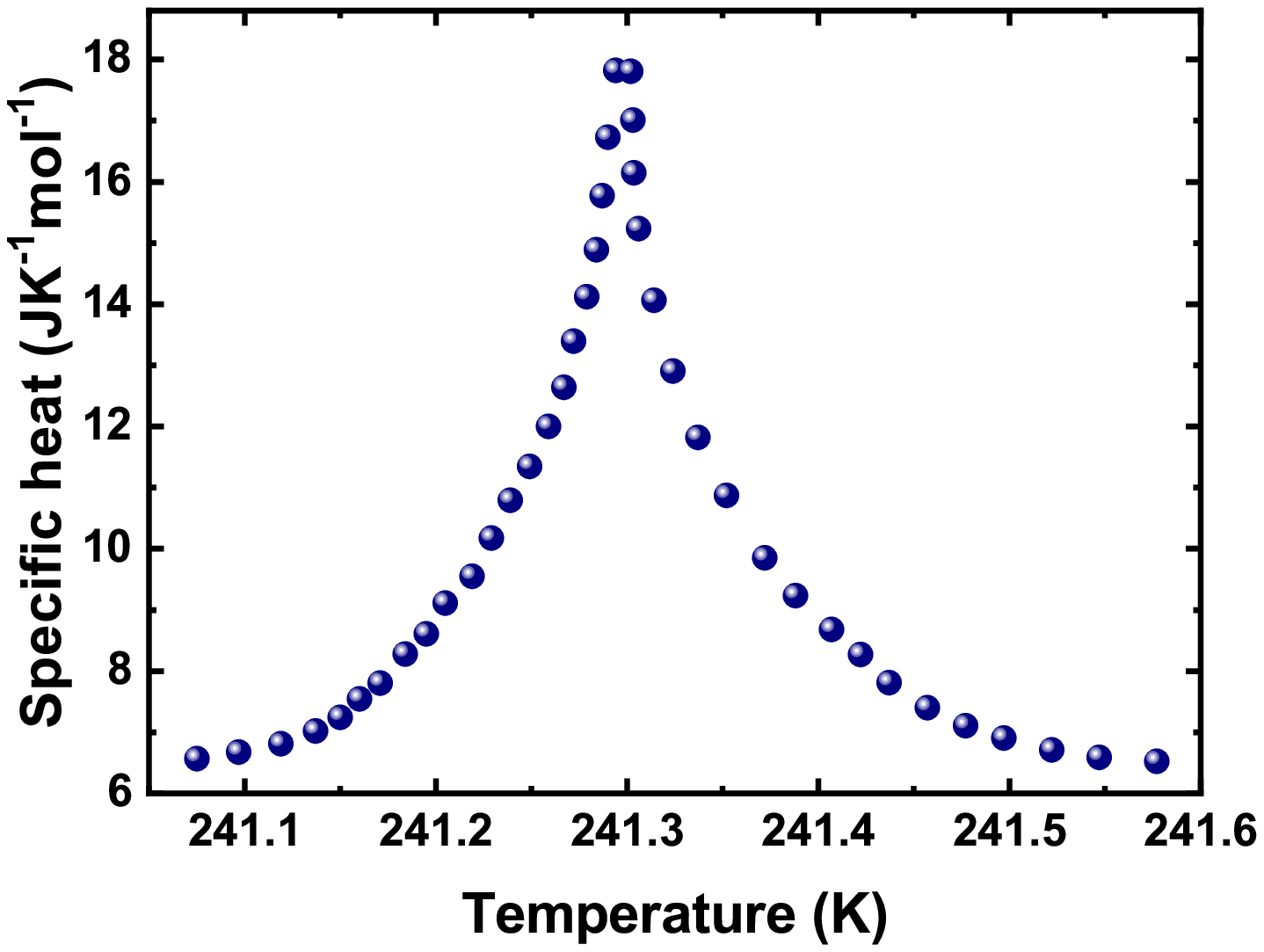}
\caption{
Specific heat of fluid electrons in $\mathrm{YH}_{9}$. The divergence of the correlation length near the critical point results in a scale-free behavior of the specific heat.
}  \label{specific1}
\end{figure}

\begin{figure}
\includegraphics[scale=0.32]{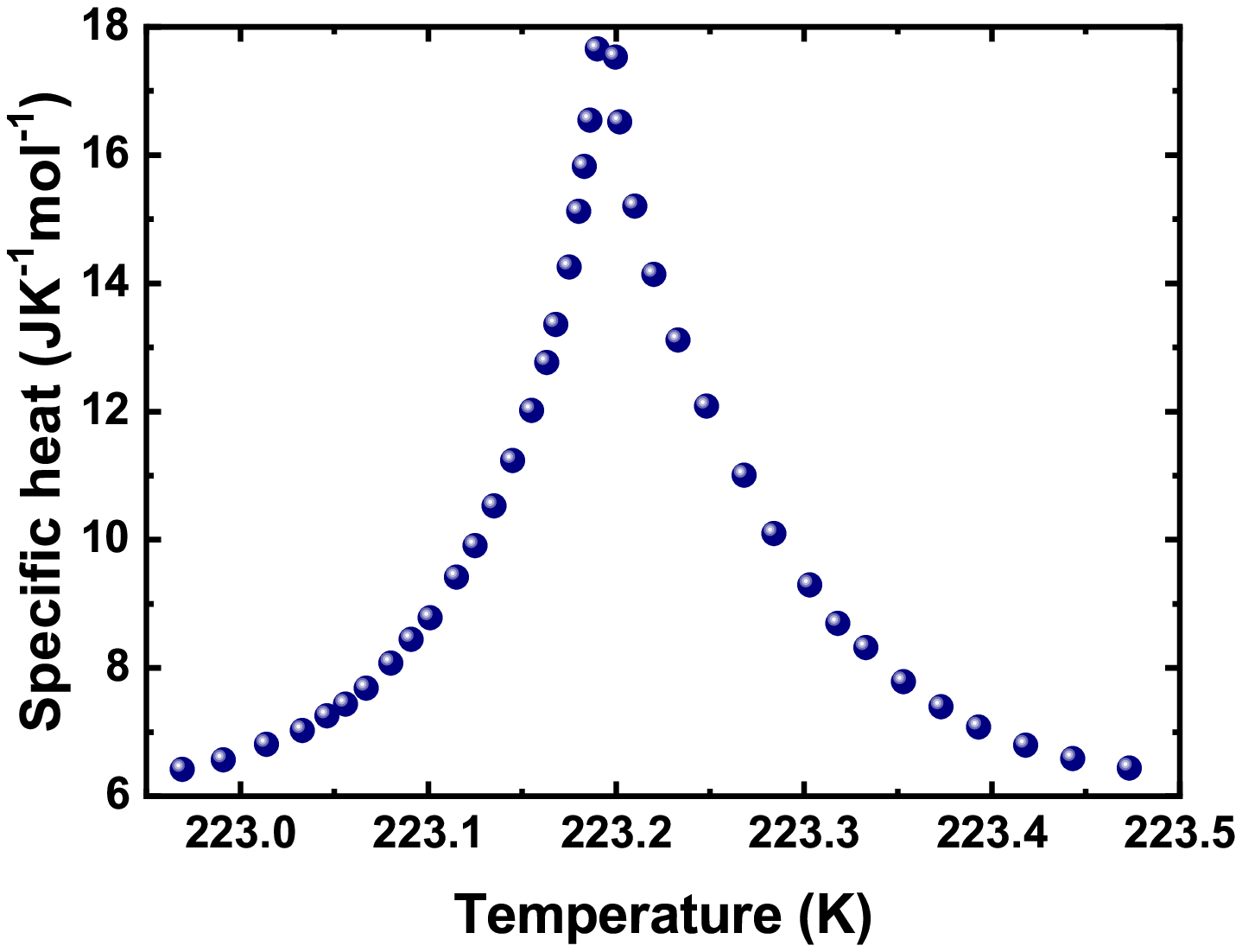}
\caption{
Specific heat of fluid electrons in $\mathrm{GdYH}_{5}$. The long range correlation length near the critical point leads to a divergence behavior of the specific heat.
}\label{specific2}
\end{figure}

An important characteristic of second-order phase transitions is the divergence of thermodynamic response functions at the critical point. This behavior reflects the system's increasing sensitivity to external perturbations as it approaches criticality.
Figures~\ref{specific1} and~\ref{specific2} illustrate the specific heat at constant volume, $C_v$, for the fluid-electron phases of $\mathrm{YH}_{9}$ and $\mathrm{GdYH}_{5}$, respectively near the highest critical temperature.
In the vicinity of the critical point, fluctuations in the free energy become increasingly significant and cannot be neglected. These fluctuations grow in magnitude as the system nears criticality, leading to the development of long-range correlations among the fluid electrons.
Such correlations originate from the coupling between e-e and e-phonon interactions, which transform initially short-range interactions into collective behavior. As a result, the system shows a divergent specific heat, indicating the critical behavior associated with a second-order phase transition.

\begin{figure}
\includegraphics[scale=0.28]{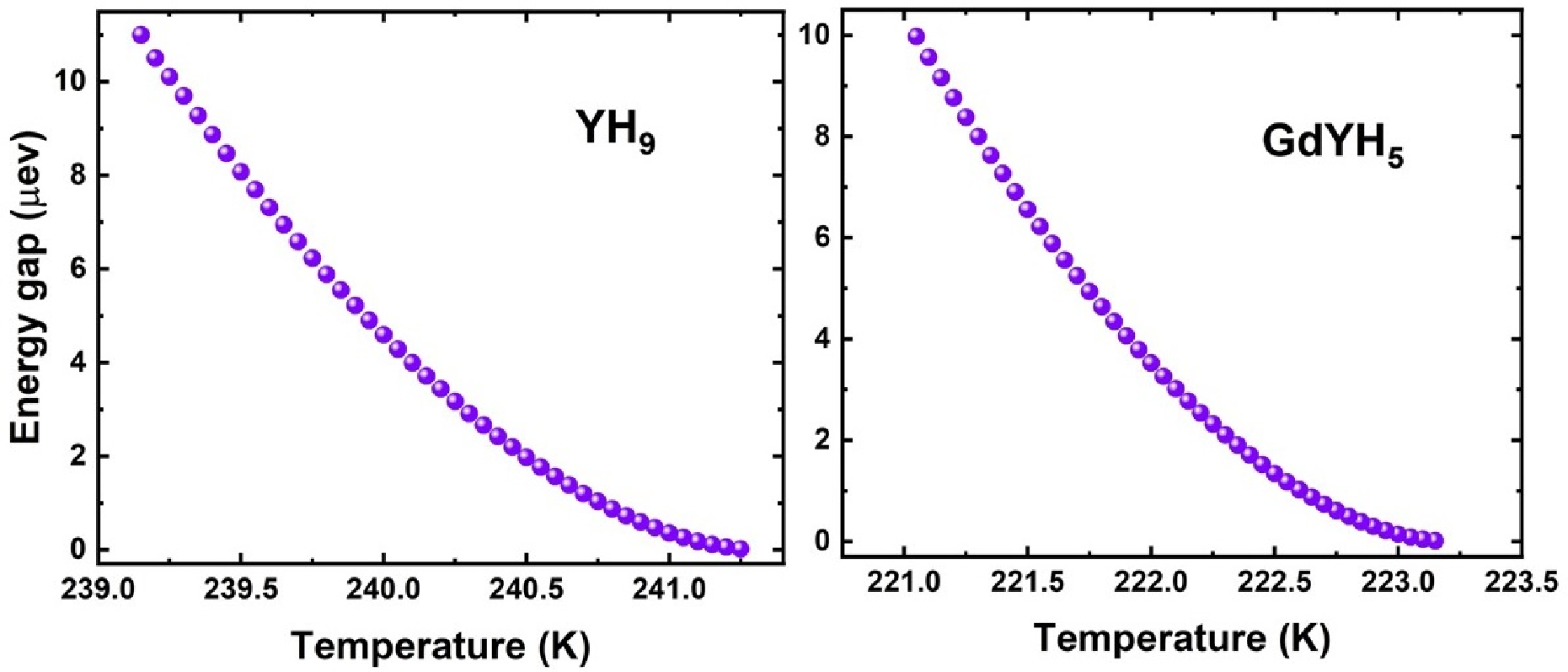}
\caption{
Temperature dependence of the superconducting energy gap $\Delta(T)$ for $\mathrm{YH}_{9}$ and $\mathrm{GdYH}_{5}$, illustrating their closure at the critical temperature $T_c$. The critical exponent $m$ and constant $d$ are obtained by fitting the curve to the superconducting energy-gap relation.
}\label{gap1}
\end{figure}

The superconducting energy gaps are determined from the total energy per particle, as formulated in Eq.~(\ref{eq2}). The computed results for $\mathrm{YH}_{9}$ and $\mathrm{GdYH}_{5}$ are displayed in Figs.~\ref{gap1}, for temperatures below the critical point.
As the temperature decreases, the magnitude of the superconducting gap $\Delta$ increases, reflecting the progressive stabilization of the Cooper-pair. This behavior is consistent with the standard superconducting energy-gap relation $\Delta \sim \left( 1-\frac{T}{T_c}\right)^m\left( 1+d\frac{T}{T_c}\right)$ \cite{arx}, confirming the reliability of the present calculations in describing the superconducting state of these hydrides. 
$d$ is a constant, while $m$ represents the critical exponent. For $\mathrm{YH}_{9}$, $d$ and $m$ take the values $1.24$ and $1.74$, respectively, whereas for $\mathrm{GdYH}_{5}$, they are $1.29$ and $1.79$.

Computing of the superconducting energy gap leads to a direct verification of the emergence of superconductivity in the new identified phase and provides information about its nature. In condensed matter physics, one of the significant issues is to distinguish between conventional and unconventional superconductors, as their microscopic mechanisms and pairing symmetries differ fundamentally.
A practical method to classify a superconductor is analyzing the dimensionless gap ratio, defined as
$\alpha = \frac{2\Delta_0}{k_B T_c},$
where $\Delta_0$ is the superconducting energy gap at zero temperature. In the BCS theory describing conventional superconductors, $\alpha$ takes a value close to $3.5$. Deviations from this value, significantly larger, indicate unconventional behavior \cite{foss}.
The calculated gap ratios are $6.27$ for $\mathrm{YH}_{9}$ and $7.06$ for $\mathrm{GdYH}_{5}$.
Such large values indicate that both compounds lie within a strongly correlated regime, deviating considerably from the weak-coupling behavior predicted by the BCS theory. Therefore, the Eliashberg formalism provides an appropriate formalism for describing the superconducting properties in this regime.

\section{Concluding Remarks}
This work reaffirms that the LOCV method is a highly applicable and reliable formalism for investigating unconventional superconductors as well. By employing the energy cluster expansion approach, the LOCV formalism enhances the precision of calculations to the point of convergence while effectively avoiding redundant computations that do not contribute to improved accuracy.
Within this framework, we have computed various thermodynamic and magnetic properties of the $\mathrm{YH}_{9}$ structure under high pressures. The excellent agreement between our LOCV results and the available experimental data of Ref.~\cite{A11} strongly supports the validity and robustness of this method. Motivated by this consistency, we further explored high-$T_c$ superconductivity in the gadolinium–yttrium–hydrogen system across different stoichiometries. Among the analyzed compounds, only $\mathrm{GdYH}_{5}$, $\mathrm{GdYH}_{6}$, $\mathrm{GdYH}_{8}$, and $\mathrm{GdYH}_{10}$, were found to show stable phases at $\sim100~\mathrm{GPa}$. Remarkably, $\mathrm{GdYH}_{5}$ demonstrates a superconducting phase transition with a critical temperature exceeding $200\ K$.

These results clearly indicate that the LOCV method provides a powerful and efficient framework for predicting stable crystal phases of novel atomic combinations and for determining their potential superconducting behaviors. This capability offers a promising path for the discovery of new superconductors without the extensive experimental sampling typically required in the laboratory.
The obtained results will be verified experimentally
An additional property of the LOCV approach lies in its computational efficiency. Unlike many other theoretical methods that require extensive additional calculations to obtain the desired results, the LOCV formalism focuses only on the essential computations, thereby minimizing both cost and computational time.
Moreover, as it directly gives the minimized free energy of the system, all relevant thermodynamic quantities can be obtained straightforwardly.
In conclusion, our findings indicate that the LOCV method not only provides accurate predictions for high-pressure superconducting hydrides but also facilitates the development of achieving high-T$_c$ superconductors at ambient pressure.

\begin{acknowledgments}
The authors sincerely thank Dr. V. Minkov for generously providing their experimental results so promptly.
\end{acknowledgments}


\end{document}